\documentclass[5p,final]{elsarticle}
\usepackage{amsfonts}
\usepackage{amssymb}
\usepackage{amsmath}  
\usepackage{amsthm}
\usepackage{enumitem} 
\usepackage{geometry}
\usepackage{graphicx}% Include figure files
\usepackage{algorithm}
\usepackage{algcompatible}
\usepackage{algpseudocode}
\usepackage{bbm}
\usepackage[colorlinks=true, allcolors=blue]{hyperref}
\usepackage{hyperref}
\usepackage[utf8]{inputenc} 
\usepackage[english]{babel} 
\usepackage{url}
\usepackage{wrapfig}
\usepackage{graphicx}
\usepackage{subfiles}
\usepackage{xfrac}
\usepackage{multirow}
\usepackage{cleveref}
\usepackage{balance}
\usepackage{svg}
\usepackage{booktabs}
\usepackage{tikz}
\usetikzlibrary{shapes.geometric, arrows}
\usepackage{tikz-3dplot}
\usepackage[squaren]{SIunits}
\usepackage{caption}
\usepackage{natbib}

\usepackage[normalem]{ulem}
\usepackage{subcaption}
\usepackage{lipsum}
\usepackage{todonotes}
\usepackage{float}
\usepackage{tocloft}
\newcommand{\bc}{\begin{center}}
\newcommand{\ec}{\end{center}}
\newcommand{\be}{\begin{equation}}
\newcommand{\ee}{\end{equation}}
\newcommand{\bea}{\begin{eqnarray}}
\newcommand{\eea}{\end{eqnarray}}
\newcommand{\beq}{\begin{eqnarray*}}
\newcommand{\eeq}{\end{eqnarray*}}
\newcommand{\bv}{\left( \begin{array}{c} }
\newcommand{\ev}{\end{array} \right) }

\newcommand{\Pp}{ \mathbb{P} }

\newcommand{\E}{\mathbb{E}}

\newcommand{\Cov}{\mathbb{C}\mathrm{ov}}

\newcommand{\N}{\mathbb{N}}

\newcommand{\sign}{\mathrm{sign}}

\usepackage{natbib}
\begin{document}
\title{Revisiting Trade-sign Long-memory and Square-root Law price impact}
\author[unsw-mathstats]{Chris Angstmann}\ead{c.angstmann@unsw.edu.au}
\address[unsw-mathstats]{School of Mathematics and Statistics, University of New South Wales, Sydney, NSW 2052, Australia}
\author[uct-sta]{Tim Gebbie} \ead{tim.gebbie@uct.ac.za}
\address[uct-sta]{Department of Statistical Sciences, University of Cape Town, Rondebosch 7701, Western Cape, South Africa}
% \email{}

\begin{abstract}
Starting with a coupled discrete reaction--diffusion formulation for the lit and latent order books with non-uniformly sampled event times and meta-order source terms we show how two familiar market-microstructure regularities can emerge from this framework: the long-memory of trade signs associated with the Lillo--Mike--Farmer (LMF) theory and the square-root law (SQRL) of meta-order impact.  This uses the locally linear order book and constant participation rate execution in the front dynamics to reduce the dynamics to a Volterra equation whose leading-order solution then yields the well known result of concave impact trajectory, and a completion impact proportional to the square root of the meta-order size. We then use the interface representation to show how heavy-tailed Pareto meta-order lengths generate power-law trade-sign autocorrelations through the source term. These are familiar derivations, what is slightly different here is that we reinterpret these known derivations to make it clear that the LMF law is an event-time sign-memory statement, whereas the square-root law is a physical-time viability statement where subordination can alter the calendar-time impact trajectories depending on the mappings and interpolation used to set continuum operational time.
\end{abstract}

\begin{keyword}
event-time \sep calendar-time \sep discrete time random walks \sep Markov lattice dynamics \\
{\it Subject Areas:} 
%\PACS 89.65.Gh \sep 02.50.Ey
\MSC 82C41 91G60 91G60 91B26
\JEL 91G70
%{\it ACM:} Finance; Electronic Trading; Data Stream Mining; Agent-Based Modeling
% https://cran.r-project.org/web/classifications/JEL.html
\end{keyword}

\maketitle
\tableofcontents

\section{Introduction} \label{sec:intro}

% I need to cite Bouchaud et al. (2004) and Bouchaud et al. (2006) and Lillo & Farmer (2004) in reference to  propagator / efficiency / criticality lillo2004efficient farmer2013impact

Reaction--diffusion and latent-order-book approaches provide a parsimonious mesoscopic description of price formation in terms of diffusive relocation, cancellation, replenishment, and local matching of opposite-side liquidity \citep{toth2011anomalous,donier2015fully,benzaquen2018multi,bouchaud2018tqp}. In parallel, the order-splitting literature initiated by \citet{lillo2005theory} explains the long memory of market-order signs through the persistence of meta-orders with heavy-tailed lengths, a relation sharpened and quantitatively validated in recent work \citep{sato2023prl,sato2024jsp}. 

The first motivation is structural. In the reaction--diffusion approach, the mid-price is a moving reaction front generated by the signed imbalance between bid and ask liquidity. In the order-splitting approach, trade-sign persistence is generated by the renewal structure of sequential meta-orders. Both phenomena are driven by the same underlying signed order flow, but the existing literature usually treats them separately: free-boundary dynamics for impact on the one hand, and renewal or hidden-order theory for sign memory on the other \citep{lillo2005theory,donier2015fully,benzaquen2018multi}. Trying to bring them together requires a formulation that keeps track of both the local state of the order book and the event-time persistence structure of meta-orders.

The modern impact literature already contains two partially overlapping explanations for concave meta-order impact.  In the latent-order-book and reaction--diffusion view, the relevant mesoscopic object is a signed liquidity field whose locally linear profile near a moving front converts diffusive liquidity transport into the nonlinear Volterra impact equation and, under constant-rate execution, the square-root completion law \citep{toth2011anomalous,donier2015fully,benzaquen2018multi,bouchaud2018tqp}.  In the propagator and empirical hidden-order literature, concavity and price diffusion are instead tied to transient or history-dependent impact, so that persistent order flow need not create predictable returns \citep{bouchaud2004fluctuations,gatheral2010no,moro2009market,bershova2013nonlinear}. 

An alternative construction by \citet{nadtochiy2022simple} does not derive concavity directly from a locally linear latent order book, but rather (equivalently) from a dynamical setting in which local impact on a fundamental-price signal is linear, while the distribution of the microstructural state changes during a VWAP-like sequence of co-directed trades -- but it is {\it a priori} a continuous-business-time model. For large-tick assets this state can be read through the microprice, and hence through best bid and ask queue volumes, where the visible-queue is best viewed as a testable derived implication rather than the primitive assumption. 

The contribution here is therefore not to re-establish the square-root law or the LMF relation in isolation, but to locate both within one discrete, non-uniform event-time bid/ask construction.  This matters because order-book data are generated by asynchronous events with irregular waiting times, whereas the standard LLOB/Volterra derivation is written after a continuum limit in an operational time that is usually identified, often implicitly, with calendar time.  Stochastic-clock and transaction-time models already show that changing the clock can change the statistical form of high-frequency returns \citep{clark1973subordinated,ane2000order}, and asynchronous-trading/Epps-effect studies show that calendar-time synchronisation can distort measured dependence in event-driven markets \citep{chang2020revisiting,gurgul2016impact}.  

Here we make this clock separation explicit: event time indexes child-order and book-update events and is the natural domain of the LMF sign process; operational time is the continuum clock of the reaction--diffusion front and Volterra equation; calendar time is the physical clock in which participation rates, execution horizons, and realised impact are measured.  In this precise sense the present formulation is more foundational with respect to the construction of time than continuous-time impact models that assume a business or calendar clock at the outset \cite{nadtochiy2022simple}.

We identify continuous-time equations as operational-time limits of an event-driven microstructure. The exact discrete imbalance reduction, the comparison of primary, source-based, and flux-based sign conventions, and the DTRW/subordination interpretation then provide the bridge between event-time sign persistence and physical-time impact transport \citep{angstmann2015dtrw,meerschaert2012fractional,diana2024anomalous,angstmann2026nonunique}.

Concretely, starting with a finite-grid, non-uniform event-time reaction--diffusion model for bid and ask liquidity, under a continuum limit this then yields the well known linear inhomogeneous reaction--diffusion equation with a moving front identified as the mid-price. We will follow the standard approach: use a local linear latent-order-book approximation near the front, then the Green-function reduction leading to the well known exact non-linear Volterra equation for the impact trajectory \cite{toth2011anomalous, donier2015fully, bouchaud2018tqp}. 

In the weak cancellation and small-displacement regime this reduces to an Abel convolution and implies transient impact proportional to $\sqrt{t}$ and completion impact proportional to $\sqrt{Q}$ at fixed participation rate. What we use this to highlight is the physical-time formulation inherent in the standard square-root law derivation. 

Independently, sequential meta-orders with heavy-tailed event-time lengths generate an exact renewal formula for the trade-sign autocorrelation and recover the LMF exponent relation $\gamma=\alpha-1$. We need to compare primary, source-based, and flux-based sign conventions, to then explain how persistent order flow in event time can coexist with weak return autocorrelation in physical time through transient or state-dependent liquidity response, and make explicit the DTRW/subordination split between event time and physical time. 

The paper proceeds as follows. \Cref{sec:model} defines the finite-grid, non-uniform event-time bid/ask reaction--diffusion model with lit and latent source terms and signed meta-order forcing and identifies the discrete reaction front. \Cref{sec:continuum} gives the formal continuum limit to a linear inhomogeneous reaction--diffusion PDE. \Cref{sec:locallinear} analyses the stationary background and demonstrates the local linear latent order book approximation near the front in the continuum representation, differentiates the front condition to obtain the free-boundary equation for the mid-price, performs the background--perturbation split, introduces the Green-function representation, and then derives the exact nonlinear Volterra equation for the front trajectory. The Volterra equation is then specialised to constant-rate execution to derive the leading-order square-root impact law in Section \ref{sec:sqrt}. \Cref{sec:tradesign} considers the trade sign process and how it is mapped into the front dynamics. Here the primary meta-order sign is compared with source-based and flux-based interface representations. An exact discrete sign-autocorrelation formula and its Pareto-tail asymptotic form are motivated to recover the LMF exponent relation. Followed by a brief discussion on how persistent order flow can coexist with weak return autocorrelation via a transient or state-dependent propagator \cite{bouchaud2018tqp}. \Cref{sec:subordination} clarifies the role of non-uniform waiting times through a DTRW/subordination perspective. 

\section{Discrete non-uniform reaction--diffusion model}
\label{sec:model}
Let the log-price axis be discretised on the finite grid with grid spacings $\Delta x > 0$ and $J \in \N$
\begin{equation}
    x_j=j\Delta x,\qquad j=0,1,\dots,J,
\end{equation}
and let child-order or order-book events occur at non-uniform physical times
\begin{equation}
    0=t_0<t_1<t_2<\cdots,
    \qquad \Delta t_m:=t_m-t_{m-1}.
\end{equation}
Here the integer index $m$ counts the event-times. The time-increments $\Delta t_m$ can be non-uniform. Trade-sign auto-correlations will typically be measured in event {\it i.e.} as a function of some lag $\ell \in \N$ whereas the impact trajectories and front motion itself can be represented in either event time or in the physical calendar time.

The bid- and ask-side order-density states immediately after event $m$ are denoted by $\rho_{B,j}^m$ and $\rho_{A,j}^m$, respectively. For each log-price node $x_j$ and event index m, $\rho_{B,j}^m\ge 0$ and $\rho_{A,j}^m \ge 0$. For interior node $j=1,\dots,J-1$ we can specify event-time updates given that between events, liquidity diffuses, cancels, is replenished by lit and latent sources, and is perturbed by signed child-order forcing. 

Now, given a survival function $\theta^{n,m-1}_i$ from event $n$ to $m-1$ and a memory kernel $K_{m-n}$, we can define a discrete time random walk model on a random lattice with diffusion intensities given in terms of left, right and self jump probabilities \cite{angstmann2015dtrw,diana2024anomalous} 
\begin{equation}
    \lambda_{j|j+i}^{m} = \tfrac{r}{2} \delta_{j-1,j+i}+ (1-r) \delta_{j,j+i}+\tfrac{r}{2} \delta_{j+1,j+i}. \label{eq:TransitionProb1}
\end{equation}
The generic symmetric bid/ask update then follows from combining the anomalous diffusions with orders that have survived, symmetric reactions across the spread, sources and meta-order forcing at the front
\begin{align}
\rho_{B,j}^{m}
=&\sum_{i=-1}^{i=+1} \lambda_{j|j+i}^{m} \sum_{n=0}^{m-1} K_{m-n} \theta_{j+i}^{n,m-1} \rho_{B,j+i}^{n}
\nonumber\\
&+ \theta^{m,m-1}_{j} \rho_{B,j}^{m-1}
-\sum_{n=0}^{m-1} K_{m-n} \theta^{n,m-1}_{j} \rho_{B,j}^{n} \nonumber \\ 
&- \kappa\Delta t_m\rho_{A,j}^{m-1}\rho_{B,j}^{m-1}
+ \Delta t_m S_{j,m}^{B}+M_{j,m}^{B},
\label{eq:rapid_bid_full}
\\
\rho_{A,j}^{m}
=& \sum_{i=-1}^{i=+1} \lambda_{j|j+i}^{m} \sum_{n=0}^{m-1} K_{m-n} \theta_{j+i}^{n,m-1} \rho_{A,j+i}^{n}
\nonumber\\
&+\theta^{m,m-1}_{j} \rho_{A,j}^{m-1}
- \sum_{n=0}^{m-1} K_{m-n} \theta^{n,m-1}_{j} \rho_{A,j}^{n} \nonumber \\
&- \kappa\Delta t_m\rho_{B,j}^{m-1}\rho_{A,j}^{m-1}
+ \Delta t_m S_{j,m}^{A}+M_{j,m}^{A},
\label{eq:rapid_ask_full}
\end{align}
where $\Delta_m$ is the waiting time between event $m-1$ and event $m$ in the operational continuous time of the front evolution. Here the sources are $S$, the meta-orders forcing the front are $M$, $\nu$ is a cancellation rate, and $\kappa$ the reaction rate.

If we assume exponential arrivals 
\begin{equation}
    \theta_j^{m,m-1} = e^{-\nu \Delta t_m},
\end{equation}
no memory 
\begin{equation}
    K_{m-n}=\delta_{n,m-1},
\end{equation} 
and that the random walk on the lattice is diffusive in the appropriate limits
\begin{equation}
    D =\lim_{\Delta x, \Delta t_m \to 0} \frac{r}{2}\frac{\Delta x^2}{\Delta t_m }.
\end{equation}
with a diffusion constant D. Then the bid/ask update has the canonical form 
\begin{align}
\rho_{B,j}^m
&= \rho_{B,j}^{m-1}
+ \frac{D\Delta t_m}{(\Delta x)^2}
\bigl(\rho_{B,j+1}^{m-1}-2\rho_{B,j}^{m-1}+\rho_{B,j-1}^{m-1}\bigr) \nonumber \\
&\quad - \nu\Delta t_m\rho_{B,j}^{m-1}
- \kappa\Delta t_m\rho_{A,j}^{m-1}\rho_{B,j}^{m-1}
\nonumber \\
&\quad + \Delta t_m S_{j,m}^{\mathrm{lit},B}
+ \Delta t_m S_{j,m}^{\mathrm{lat},B}
+ M_{j,m}^{B},
\label{eq:discB}
\\[1ex]
\rho_{A,j}^m
&= \rho_{A,j}^{m-1}
+ \frac{D\Delta t_m}{(\Delta x)^2}
\bigl(\rho_{A,j+1}^{m-1}-2\rho_{A,j}^{m-1}+\rho_{A,j-1}^{m-1}\bigr)
\nonumber\\
&\quad - \nu\Delta t_m\rho_{A,j}^{m-1}
 - \kappa\Delta t_m\rho_{A,j}^{m-1}\rho_{B,j}^{m-1}
\nonumber\\
&\quad
+ \Delta t_m S_{j,m}^{\mathrm{lit},A}
+ \Delta t_m S_{j,m}^{\mathrm{lat},A}
+ M_{j,m}^{A}.
\label{eq:discA}
\end{align}
Here $D>0$ is the diffusion coefficient, $\nu\ge 0$ the cancellation rate, and $\kappa\ge 0$ the local matching rate. Boundary values are explicitly given at $j=0$ and $j=J$. We have also implicitly identified the operational time of the front with the calendar time of the system that measure the front dynamics, some physical time $t$. 

Now, let $p_{m-1}$ denote the pre-update front estimate and define the signed distance
\begin{equation}
    \xi_j^{m-1}:=x_j-p_{m-1}.
\end{equation}
The lit source terms are exponentially concentrated near the front and the latent source terms vanish at the front and approach non-zero far-field levels.

The lit market source kernels in \cref{eq:discB,eq:discA} are
\begin{align}
S_{j,m}^{\mathrm{lit},B}
&=
\Lambda_\ell^B\exp\!\left(\frac{\xi_j^{m-1}}{\ell_\ell^B}\right)\mathbbm 1_{\{\xi_j^{m-1}\le 0\}},
\\
S_{j,m}^{\mathrm{lit},A}
&=
\Lambda_\ell^A\exp\!\left(-\frac{\xi_j^{m-1}}{\ell_\ell^A}\right)\mathbbm 1_{\{\xi_j^{m-1}\ge 0\}},
\end{align}
where the lit-order source intensities are: $\Lambda_\ell^B, \Lambda_\ell^A\ge 0$, and with respective length scales: $\ell_\ell^B,\ell_\ell^A \ge 0$. These govern the source terms near the reaction front. 

The latent order book source kernels in \cref{eq:discB,eq:discA} have latent-order intensities $\Lambda_u^B, \Lambda_u^A\ge 0$, and length scales: $\ell_u^B,\ell_u^A \ge 0$ to then govern the non-zero far-field order reservoirs
\begin{align}
S_{j,m}^{\mathrm{lat},B}
&=
\Lambda_u^B\Bigl[1-\exp\!\left(\frac{\xi_j^{m-1}}{\ell_u^B}\right)\Bigr]\mathbbm 1_{\{\xi_j^{m-1}\le 0\}},
\\
S_{j,m}^{\mathrm{lat},A}
&=
\Lambda_u^A\Bigl[1-\exp\!\left(-\frac{\xi_j^{m-1}}{\ell_u^A}\right)\Bigr]\mathbbm 1_{\{\xi_j^{m-1}\ge 0\}}.
\end{align}

The side-specific forcing terms are represented as point impulses on the grid with rates $\mu_q^B$ and $\mu_q^A$ ,
\begin{align}
M_{j,m}^{B}&=\sum_q \mu_q^B\delta_{m,m_q}\delta_{j,j_q}, \\
M_{j,m}^{A}&=\sum_q  \mu_q^A\delta_{m,m_q}\delta_{j,j_q}.
\end{align}
Here meta-orders are represented by a sequence of child-order impulses with signs denoted by $\sigma \in \{ +1,-1\}$. The meta-orders have length $L \in \N$ and total signed volume $Q$ such that for a constant-rate execution over a horizon $T$ the signed execution rate is $m_0$ so that $Q=m_0 T$. The trade sign process can then be defined as
\begin{equation}
    \epsilon_m = \sigma_m \in\{ +1,-1\}.
\end{equation}

\subsection{Imbalance reduction}
\label{sec:imbalance}
Define the imbalance field by
\begin{equation}
    \phi_j^m:=\rho_{B,j}^m-\rho_{A,j}^m.
\end{equation}
Here a positive imbalance means bid density exceeds ask density. Subtracting \cref{eq:discA} from \cref{eq:discB} gives the exact discrete imbalance equation
\begin{align}
    \phi_j^m
    &=\phi_j^{m-1}
    +\frac{D\Delta t_m}{(\Delta x)^2}\bigl(\phi_{j+1}^{m-1}-2\phi_j^{m-1}+\phi_{j-1}^{m-1}\bigr)
  \nonumber \\ &\quad -
    \nu\Delta t_m\phi_j^{m-1}
    +
    \Delta t_m s_{j,m}
    +
    \mathcal M_{j,m},
\label{eq:discimb}
\end{align}
where the source terms for the order book are
\begin{equation}
s_{j,m}:=S_{j,m}^{\mathrm{lit},B}-S_{j,m}^{\mathrm{lit},A}+S_{j,m}^{\mathrm{lat},B}-S_{j,m}^{\mathrm{lat},A},
\end{equation}
and those for the forcing terms
\begin{equation}
    \mathcal M_{j,m}:=M_{j,m}^{B}-M_{j,m}^{A}.
\end{equation}
The term proportional to the local matching rate $\kappa$ cancels exactly because it is symmetric in bid and ask densities. This cancellation is the simplification that makes the later front-dynamics derivation trivially tractable. Each child order event acts near the front with a net imbalance
\begin{equation}
    \mathcal M_{j,m} = \sigma_m q_{j,m}
\end{equation}
where $\sigma_m \in \{ +1,-1\}$, the order $q_{j,m}\ge0$ for all grid points $j$, and $q_{j,m}$ is supported inside a small neighbourhood of the front. 

This is a key simplification of the model because the original bid/ask system is non-linear because of the local matching term $-\kappa \rho_{A,j}^{m-1} \rho_{B,j}^{m-1}$. After reduction this nonlinearity disappears from the imbalance dynamics. The remaining equation is then linear in $\phi$ but retains inhomogeneity through the net source $s_{j,m}$ and the meta-order forcing term $\mathcal M_{j,m}$. Although the signed imbalance field preserves the signed asymmetry between bid and ask liquidity, and therefore remains the natural variable for front dynamics and impact, imbalance alone no longer determines the separate bid from ask densities without additional information. The reduction is exact for $\phi$ but not invertible without additional structure. 

At this junction the trade-sign process is still defined either from the sign of the active meta-order (and not strictly from the sign of the forcing function $\mathcal M_{j,m}$ or from the imbalance flux at the reaction front). What is important to realise here is that this means that the sign of the forcing enters only through  the meta-order sign \footnote{Although the trade signs are here defined directly in terms of the meta-order signs, there are in fact two natural interface representations: the source representation (from the sign of the effective forcing $\mathcal M_{j,m}$ near the front), and the flux representation (from sign of the flux imbalance through the front). We will use the source representation and then show equivalence under event based one-step dominance and monotone flux response (Section \ref{sec:signs})}, while the magnitude of the order profile $q_{j,m}$ is non-negative and localised near the front. This means that for every child-order event for which the aggregate forcing near the front is non-zero
\begin{equation}
    \epsilon_m = \sigma_m.
\end{equation}
It must be noted that if the child-order acts far from the front, the meta-order sign may still be well defined while the source-based sign becomes weak or ambiguous. 

\subsection{Front identification}
\label{sec:front}
The discrete front location can be found after an update to event $m$ where the front price grid index $j_m^*$ is such that $\phi^m_{j_m^*} \phi^m_{j_m^*+1} \le 0$ and then the position of the front can be found using interpolation
\begin{equation}
    p_m = x_{j_m^*} + \Delta x \left( {\frac{-\phi_{j_m^*}^m}{\phi_{j_m^*+1}^m-\phi_{j_m^*}^m}} \right).
\end{equation}
This gives the mid-price at event $m$ at price grid index $j_m^*$ such that $\phi(p_m,t_m)\approx 0$.

\section{Continuum limit}
\label{sec:continuum}

Divide \cref{eq:discimb} by $\Delta t_m$ to obtain
\begin{align}
\frac{\phi_j^m-\phi_j^{m-1}}{\Delta t_m}
&=
D\frac{\phi_{j+1}^{m-1}-2\phi_j^{m-1}+\phi_{j-1}^{m-1}}{(\Delta x)^2}
\nonumber \\
&\quad
- \nu\phi_j^{m-1}
+s_{j,m}
+ \frac{\mathcal M_{j,m}}{\Delta t_m}.
\end{align}
Assume smooth interpolation $\phi^{\Delta x,\Delta t}(x_j,t_m) = \phi_j^m$ for $j =0,1,\ldots,J$ and convergence of the source $s^{\Delta x,\Delta t}(x,t) \to s(x,t)$ and forcing interpolants $m^{\Delta x,\Delta t}(x,t) \to m(x,t)$ \footnote{Here the forcing interpolant is $m^{\Delta x,\Delta t}(x_j,t_m) = \frac{\mathcal M_{j,m}}{\Delta t_m}$}. Now we use that the maximal event-time increment tends to zero {\it i.e} $ \sup_{m} \Delta t_m \to 0$, and that the interpolated imbalance satisfies
\begin{equation}
    \frac{\phi_j^m-\phi_j^{m-1}}{\Delta t_m} \to \partial_t \phi(x_j,t)
\end{equation}
in a pointwise or weak sense which is sufficient for the formal passage to the limit. We also need to assume spatial consistency by assuming that as $\Delta x \to 0$ with some $J \Delta x \to L_x$, as we are on a finite grid on $x \in [0, L_x]$. We have that
\begin{equation}
    \frac{\phi_{j+1}^{m-1}-2\phi_j^{m-1}+\phi_{j-1}^{m-1}}{(\Delta x)^2} \to \partial_{xx} \phi(x_j,t).
\end{equation}
Then with sufficient diffusion stability
\begin{equation}
    \max_m \frac{D \Delta t_m}{\Delta x^2} \le \frac{1}{2},
\end{equation}
one then recovers the continuum limit 
 \begin{align}
    \partial_t\phi(x,t)
    &=D\partial_{xx}\phi(x,t)
    -\nu\phi(x,t) \nonumber \\
    \quad &
    +s^{\mathrm{lit}}(x,p(t))
    +s^{\mathrm{lat}}(x,p(t))
    +m(x,t).
\label{eq:pde}
\end{align}
The continuum equation remains linear in the imbalance field (as in the discrete setting) and now all the non-trivial structure enters through: source terms for the moving front $p(t)$, the external meta-order forcing term $m(x,t)$, and the boundary conditions inherited from the finite grid formulation. This linear structure will allow us to use a Green function representation and to then derive an explicit front equation and impact equation \cite{toth2011anomalous,donier2015fully,bouchaud2018tqp}.

\subsection{Continuum sources}

In the continuum limit, the corresponding formal source fields are written relative to the front position $p(t)$
\begin{equation}
    s(x,t)=s^{\mathrm{lit}}(x,p(t))+s^{\mathrm{lat}}(x,p(t)).
\end{equation}
Here the sources are
%\begin{widetext}
\begin{align}
    s^{\mathrm{lit}}(x,p)
    &=\Lambda_\ell^B\exp\!\left(\frac{x-p}{\ell_\ell^B}\right)\theta(p-x)
    \nonumber\\
    &\quad -
    \Lambda_\ell^A\exp\!\left(-\frac{x-p}{\ell_\ell^A}\right)\theta(x-p),
    \label{eq:stage3_lit_cont}
    \\
    s^{\mathrm{lat}}(x,p)
    &=
    \Lambda_u^B\Bigl[1-\exp\!\left(\frac{x-p}{\ell_u^B}\right)\Bigr]\theta(p-x) \nonumber\\
    &\quad -
    \Lambda_u^A\Bigl[1-\exp\!\left(-\frac{x-p}{\ell_u^A}\right)\Bigr]\theta(x-p),
    \label{eq:stage3_lat_cont}
\end{align}
%\end{widetext}
where $\theta$ denotes the Heaviside step function.

The continuum forcing term $m(x,t)$ is the physical-time analogue of the discrete net imbalance forcing $\mathcal M_{j,m}$. This term will be specialised either to a localised source concentrated near the front, or to a simpler effective forcing used in impact calculations.

\subsection{Boundary conditions}

The discrete model is posed on a finite grid, the continuum limit initially inherits a finite spatial interval and boundary data. Formally, if $\phi_0^m\to \phi_L(t)$, and $\phi_J^m\to \phi_R(t)$, as $\Delta x\to0$, then the PDE is posed on $x\in[0,L_x]$ with boundary conditions
$\phi(0,t)=\phi_L(t)$, and $\phi(L_x,t)=\phi_R(t)$. An unbounded-domain idealisation would then correspond to replacing these explicit finite-domain boundary conditions by appropriate asymptotic conditions.

\section{Local-linear order book} \label{sec:locallinear}

We are now ready to analyze the stationary version of the continuum imbalance and to isolate the local structure of the imbalance near the reaction front. The stationary state represents the background order-book configuration about which later moving-front or impact calculations will be considered. This is the baseline shape of the imbalance maintained by diffusion, cancellation and replenishment {\it i.e.} we have $\phi(x,t) = \phi^*(x)$, and $p(t) = p_0$ for a fixed front location $p_0$ with no active meta-order forcing: $m(x,t)=0$. 

Setting $\partial_t\phi=0$ and $m=0$ in \cref{eq:pde} yields the stationary imbalance profile $\phi^*(x)$ at the front position $p_0$
\begin{equation}
D\,\frac{d^2\phi^*}{dx^2}(x)-\nu\phi^*(x)+s^{\mathrm{lit}}(x,p_0)+s^{\mathrm{lat}}(x,p_0)=0, \label{eq:stationary}
\end{equation}
where the front condition is $\phi^*(p_0)=0$. Taylor expanding around $x=p_0$ gives
\begin{align}
\phi^*(x)&=\phi^*(p_0)+\partial_x\phi^*(p_0)(x-p_0)\nonumber\\
&\quad +\tfrac12\partial_{xx}\phi^*(p_0)(x-p_0)^2+{\cal O}((x-p_0)^3).
\end{align}
Using $\phi^*(p_0)=0$ and defining $\mathcal L=-\partial_x\phi^*(p_0)>0$\footnote{The stationary front is a simple zero {\it i.e.} $\partial_x \phi^*(p_0) \neq 0$ (transversality). If the zero were degenerate the local inversion required to convert changes in the imbalance into changes in the front would fail or become singular. This is a crucial conditions for the free-boundary derivations.}, the local liquidity slope, yields
\begin{equation}
\phi^*(x)=-\mathcal L(x-p_0)+\frac12\partial_{xx}\phi^*(p_0)(x-p_0)^2+{\cal O}((x-p_0)^3).
\end{equation}
The leading term is the Local Linear Order Book (LLOB) approximation $\phi^*(x)= -\mathcal L(x-p_0)+{\cal O}\bigl((x-p_0)^2\bigr)$. The stationary problem is setup on the finite interval $x \in [0,L_x]$ with the boundary conditions inherited from the discrete model $\phi^*(0)= \phi_L$ and $\phi^*(L_x)=\phi_R$. However, the local front dynamics only depend on the behaviour of the stationary profile in the immediate neighbourhood of $x=p_0$. This means that if the profile is sufficiently smooth near the front, then the front is well separated from the boundaries; the finite interval problem can then be approximated locally by what is for all practical purposes an unbounded domain. That being said, one must be clear that this does not replace the finite-grid origin of the model globally.

\subsection{Front dynamics} \label{ssec:frontvel}

The front is a free-boundary and it is implicitly defined by the zero of the imbalance field. Differentiate the front condition $\phi(p(t),t)=0$ using the chain rule:
\begin{equation}
0=\frac{d}{dt}\phi(p(t),t)=\partial_t\phi(p(t),t)+\dot p(t)\partial_x\phi(p(t),t).
\end{equation}
Solving for $\dot p(t)$ gives
\begin{equation}
\dot p(t)=-\frac{\partial_t\phi(p(t),t)}{\partial_x\phi(p(t),t)}.
\end{equation}
Substitute the continuum PDE and use $\phi(p(t),t)=0$ to eliminate the cancellation term $-\nu \phi$ at the front. This yields the exact free-boundary equation
\begin{equation}
    \dot p(t)= -\frac{D\partial_{xx}\phi+s^{\mathrm{lit}}+s^{\mathrm{lat}}+m}{\partial_x\phi}\Bigg|_{x=p(t)}.
\label{eq:frontvel}
\end{equation}
The front velocity equation separates the drivers of front motion in four terms: diffusive curvature that captures the local curvature of the imbalance profile and hence the tendency for diffusion to reshape the front neighbourhood, lit order book replenishment from passive liquidity injected near the front, latent order replenishment from long term investors as it becomes available near the front, and the meta-order forcing associated with injections of signed order-flow from external immediate execution flow. These are all filtered by the inverse slope $1/\partial_x \phi(p(t),t)$ which converts local imbalance into front displacement. Under the local linear approximation the inverse slope is approximated: $\partial_x \phi(p(t),t) \approx -{\cal L}$ for some ${\cal L}>0$.

The local linear approximation provides a link between the imbalance perturbations and the front displacement. If the stationary profile is perturbed by a small signed local imbalance increment $\delta q$ (say) near the front. Then to leading order $0=\phi^*(p_0+\delta p) - \delta q \approx - {\cal L} \delta p + \delta q$ which implies that $\delta p \approx \tfrac{\delta q}{\cal L}$. 

This is a local infinitesimal response formula and it does not yet describe cumulative impact over a sequence of trades. Repeated forcing is filtered through diffusion and replenishment over time. It is important to realise that there is no contradiction between local linear response and a globally concave meta-order impact.  

It is also important to notice that the local linear approximation does not require full bid/ask symmetry; it only requires a smooth stationary profile with a simple zero at the front. It becomes convenient to shift the front to the origin and set up the source structure symmetric across the front. The use of symmetry is a convenience and not a consequence of the local linear approximation. 

Equation \ref{eq:frontvel} is not yet a closed ordinary differential equation for $p(t)$ in isolation because the field $\phi(x,t)$ still depends on the full PDE solution. However, the equation becomes more tractable with a decomposition into perturbations around the stationary background, a Green-function representation of the execution induced perturbations, and then formulating this as a non-linear Volterra equation for the front trajectory in the locally linear regime -- this is well known in the literature. 

\subsection{Background-perturbation decomposition} \label{ssec:background}

We write the imbalance field in terms of the stationary field $\phi^*$ from equation \ref{eq:stationary}
\begin{equation}
    \phi(x,t) = \phi^*(x) + \psi(x,t)
\end{equation}
where $\psi(x,t)$ is now the perturbation induced by the meta-order and by the subsequent front motion. We will choose a reference frame such that $p_0=0$ placing the front at the origin in the neighbourhood relevant for execution with $\phi^*(x) = - {\cal L}x$. All the perturbations will be in the execution regime where the execution regime is short enough that the front is local enough and the stationary structure can be frozen into the background: $s(x,p(t))-s(x,p_0) \approx 0$. Allow the meta-order forcing to only act at the moving boundary: $m(x,t) = m(t) \delta(x-p(t))$ for the signed execution rate $m(t)$. It is here that we are explicitly simplifying a general finite-domain continuum model to the standard locally linear impact regime known in the literature. The perturbation equation becomes
\begin{equation}
    \partial_t \psi(x,t) = D \partial_{xx} \psi(x,t) - \nu \psi (x,t) + m(t) \delta (x- p(t)), \label{eq:perturbation}
\end{equation}
with the initial conditions $\psi(x,0)=0$.
\subsection{Green-function representation} \label{ssec:Green}

If we assume that during the execution horizon that the front remains sufficiently far from the boundaries that boundary reflections are negligible on the timescales of interest. Then the finite-interval Green function can be replaced to leading order with the full-line Green function. 

In summary: we write $\phi=\phi^*+\psi$ and subtract the stationary equation from the full PDE:
$\partial_t\psi=
D\partial_{xx}\psi-\nu\psi$. 
Freeze the source differences into the background and localising the forcing {\it i.e.} from Equation \ref{eq:perturbation} with the full-line Green function
\begin{equation}
G_\nu(x,t)=\frac{e^{-\nu t}}{\sqrt{4\pi Dt}}\exp\!\left(-\frac{x^2}{4Dt}\right)\mathbbm 1_{\{t>0\}},
\end{equation}
Duhamel's principle gives \cite{donier2015fully}
\begin{equation}
\psi(x,t)=\int_0^t m(s)G_\nu(x-p(s),t-s)\,ds.
\end{equation}
Evaluating at $x=p(t)$ and using $\phi(p(t),t)= -\mathcal L p(t)+\psi(p(t),t)=0$ yields the exact nonlinear Volterra equation \cite{donier2015fully}
\begin{equation}
    \mathcal L p(t)=\int_0^t m(s)
    \frac{e^{-\nu(t-s)}}{\sqrt{4\pi D(t-s)}}
    \exp\!\left(-\frac{(p(t)-p(s))^2}{4D(t-s)}\right)
    ds.
\label{eq:volterra}
\end{equation}
This is the standard LLOB Volterra formulation of meta-order impact \cite{donier2015fully,toth2011anomalous}. The model is important because it makes explicit why impact dynamics are non-local in time. Each child order increment $m(s) ds$ contributes to the current front through a diffusion kernel proportional to $(t-s)^{-\frac{1}{2}}$. The exponential kernel is a self-correction encoding that the source was injected at the earlier front position $p(s)$ not at the current front position $p(t)$. The damping factor suppresses older order flow when cancellations are present, and the liquidity slope $\cal L$ converts the cumulative perturbation of the imbalance into price displacements. The equation is exact within the locally linear execution regime. The literature highlights two special regimes \cite{toth2011anomalous,donier2015fully}: the weak cancellation regime (on short-horizons), and the small-displacement regime (when the front displacement is small relative to the diffusion scale). 

The Short-horizon regime $e^{-\nu(t-s)} \approx 1$ allows the equation to be reduced to an undamped heat-kernel --- this is often used for the leading order approximation of the square-root law. However, it is the small-displacement regime that is of interest here because this will allow the exponential to be linearised to first order. 

This will produce the Abel-type convolution kernel that provides the canonical derivation of the square-root law. To do this we now specialise the Volterra equation in the standard way to the constant-rate executions and then exploit the small-displacement regime to derive the well known transient impact law $p(t) \propto \sqrt{t}$ and this will then give us the completion impact law $I(Q) \propto \sqrt{Q}$ at a fixed participation rate \cite{bouchaud2009markets,toth2011anomalous,donier2015fully}. The damped kernel will of course reduce to the standard heat-kernel form when the weak cancellation approximation is used. This is done to explicitly make clear the physical time dependency inherent in the square-root law.  

\subsection{Square-root impact law} \label{sec:sqrt}

% cite Moro et al. (2009) on hidden orders, explicitly reports strongly concave, approximately square-root impact and post-trade decay. 

Under weak cancellation and small displacement,
\begin{equation}
\mathcal L p(t)\approx \int_0^t \frac{m(s)}{\sqrt{4\pi D(t-s)}}\,ds.
\end{equation}
For $m(s)=m_0\mathbbm 1_{[0,T]}(s)$ and $0\le t\le T$,
\begin{equation}
\mathcal L p(t)\approx m_0\int_0^t \frac{ds}{\sqrt{4\pi D(t-s)}}.
\end{equation}
Use the substitution $u=t-s$ to obtain
\begin{equation}
\int_0^t \frac{ds}{\sqrt{4\pi D(t-s)}}
=
\frac{1}{\sqrt{4\pi D}}\int_0^t u^{-1/2}du
=
\sqrt{\frac{t}{\pi D}}.
\end{equation}
Hence the transient square-root in time law
\begin{equation}
 p(t)\approx \frac{m_0}{\mathcal L}\sqrt{\frac{t}{\pi D}}, \quad 0 \le t \le T.
\end{equation}
At completion $Q=m_0T$, so
\begin{equation}
I(Q,T)=p(T)\approx \frac{1}{\mathcal L\sqrt{\pi D}}\sqrt{m_0Q}.
\end{equation}
Substituting $m_0=\eta J$ gives the fixed-participation transient square-root law given that $J>0$ is a characteristic background flux and participation rate $\eta \in (0, 1]$
\begin{equation}
    I(Q) \approx \sqrt{\frac{\eta J}{\pi D {\cal L}^2}} \sqrt{Q}.
\end{equation}
This arises from three key ingredients: the local linear liquidity (from slope $\cal L$), diffusive time propagation (spread by the kernel $(t-s)^{-\frac{1}{2}}$), and the volume-time relation that sets a fixed execution rate so that the execution duration is proportional to the total volume. Two key points can be made; first, on scaling, and second on the role of cancellation. 

Scaling: What is important to notice here is that under a self-similar ansatz (say) $p(t) = A \sqrt{Dt}$ substitution into the Volterra equation shows that the nonlinear exponential changes the prefactor A but will preserve the $\sqrt{t}$ calendar time dependence, and this will preserve the $\sqrt{Q}$ completion impact scaling at a fixed participation rate.  

Cancellations: If cancellation is not negligible then the kernel acquires the damping factor $e^{-\nu(t-s)}$. For short horizons this will only change the prefactor, for long horizons close the characteristic scale $\nu^{-1}$ the crossover can modify the transient impact and then the pure square-root need not hold. This means that the weak cancellation assumption is necessary here, at least if the leading order square-root law is invoked. 

\section{Trade-sign process} \label{sec:tradesign}

In a bid/ask density model, a buy market order would reduce ask order density $\rho_A$, increasing the imbalance $\rho_B-\rho_A$, to produce positive imbalance forcing $\mathcal M$. Similarly, a sell market order would consume bid liquidity, reducing $\rho_B$, so producing negative $\mathcal M$ -- this is consistent with the sign of the meta order because a buyer would have $\mathcal M >0$ and a seller $\mathcal M<0$. Hence, here in imbalance variables, a buy market order is represented as a positive impulse because it removes ask liquidity or equivalently increases the imbalance; and a sell market order is represented as a negative impulse.

Now, there are three distinct sign choices arising from three distinct random processes for these impulses: a primary sign convention that follows directly from the sign of the meta order ($\epsilon_m=\sigma_m$), a sourced-based interface sign tied to the sign of the net forcing $\epsilon_m = \sign({\cal M}_{j,m})$ from ${\cal M}_{j,m} = {\cal M}_{j,m}^B - {\cal M}_{j,m}^A$ near the reaction front, and last, a flux-based interface sign $\epsilon = \sign(J_{\phi})$ tied to the sign of event-induced imbalance flux $J_{\phi} = - D \partial_x \phi(x,t)$ through the front in terms of the operational time of the model. These can have different autocorrelations and different long-memory exponents. 

This is important because the LMF Theory is an {\it event-time} object.  Given a fixed sign convention the trade-sign autocorrelation function is then 
\begin{equation}
    C_{\tau}(\epsilon) = \E [\epsilon_m \epsilon_{m+\tau}], \quad \tau \in \N.
\end{equation}
These signs are only equivalent if there is both one-step force domination near the front, and monotone flux-response and the continuum operational time and physical time conform and are measured by the parameter $t$. In general $C_{\tau}(\epsilon) \neq C_{\tau}(\epsilon^{\cal M}) \neq C_{\tau}(\epsilon^F)$. 

If over the duration of a single child-order event the increment in the imbalance field near the front is dominated by the forcing term so that one-step contributions from diffusion, cancellation and replenishment are lower order in the sign determination of the local front response then to leading order the primary sign and net forcing sign coincide.

The mapping from the forcing to the flux increment is more subtle and depends on the local order book shape, the time resolution, and the background source terms. This means that this is essentially a local structural assumption rather than a consistency claim. In the local non-degenerate regime near the front, the sign of the event-induced flux increment (say) $\Delta F_m$ is taken to be monotonic in the sign of the localised forcing. This is equivalent for both a positive local forcing near the front increasing the front flux increment, and a negative local forcing decreasing it. 

\subsection{Source-based sign representation}

Concretely, the pointwise source-based interface sign is
\begin{equation}
    \epsilon^{\cal M}_m= \sign({\cal M}_{j^*_m,m})
\end{equation}
for ${\cal M}_{j^*_m,m}\neq 0$. To allow for forcing spread over a small neighbourhood of the front, rather than at a single grid point, we have a local ball of radius $r^*$ near the front, the set of points in this ball are: ${\cal K}_m = \{ j: |x_j - p_m | \le r^* \}$. This then gives an aggregate or average front forcing as $ \bar {\cal M}_{m} = \sum_{j \in {{\cal K}_m}}{\cal M}_{j,m}$. This then allows one to define the local-window source-based sign as $\epsilon_m^{\cal M}= 
\sign (\bar {\cal M}_{m})$ for $\bar {\cal M}_{m} \neq 0$.

\subsection{Flux-based sign representation}

The event-induced change in the imbalance flux through the front can be used to determine the trade sign, but the raw flux at the front can be non-zero even in a stationary book and so does not itself naturally encode trade direction. For a discrete imbalance field the centered front flux at a child-order event is
\begin{equation}
    F_m = - \frac{1}{2} \frac{D}{\Delta x} \left( {\phi^m_{j^*_m+1}-\phi^m_{j^*_m-1}} \right).
\end{equation}
Now, if the front flux immediately before the $m$-th child-order is $F_m^-$ and the front flux immediately after this event is $F_m^+$, then the event-induced front-flux increment is $\Delta F_m = F_m^+ -F_m^-$ and the discrete flux-based sign is then 
\begin{equation}
    \epsilon_m^F =\sign(\Delta F_m).
\end{equation}
This is an interface-response sign, and not necessarily a return or price fluctuation sign: $\sign (p_m - p_{m-1})$.

In the continuum representation the imbalance flux is $J_{\phi} =  - D \partial_x \phi(x,t)$ so that for times immediately before $t_m^-$ and after $t_m^+$ the $m$-th child order event we can define the front flux increment as $\Delta J_m = J_{\phi}(p(t_m^+),t_m^+) - J_{\phi}(p(t_m^-),t_m^-)$ to give the flux-based sign as $\epsilon^F_m = \sign(\Delta J_m)$. Here the slope of the raw front flux mainly reflects the slope orientation of the order book and can remain fixed while buy and sell meta-orders alternate --- the sign that is relevant for order flow is the change in the front flux induced by the event and not the background flux. 

It should be noticed that $\sign(\Delta F_m) = \sigma_m$ only holds under the local monotone, forcing-dominated assumption; otherwise no sign-preservation is guaranteed. A mismatch can occur for several reasons. Diffusion smoothing can occur when a  local impulse affects gradients after spatial spreading, but not necessarily instantaneously at the front. Front movement itself can introduces a moving-coordinate effect. Curvature {\it i.e.} if the stationary profile is not exactly linear, then moving the evaluation point changes the measured gradient. Cancellation/replenishment source terms can offset the impulse. Discretisation can lead to the front jumping across grid nodes then {\it e.g.} a different finite element implementation may measure a different local slope. Lastly, timestamping from the mapping of operational time to event time can cause flux increments to occur over a different resolution than trade signs.

\subsection{Interface representation}  \label{sec:signs}

The primary trade-sign process is defined in child-order event time by $\epsilon_m=\sigma_m$, the sign of the active meta-order. We define a source-based sign from the forcing near the front and a flux-based sign from the event-induced change in front flux. Under signed localised forcing of the form
\begin{equation}
\mathcal M_{j,m}=\sigma_m q_{j,m},\qquad q_{j,m}\ge 0,
\end{equation}
with front-window aggregate $\mathcal M_m^{*}=\sum_{j\in\mathcal K_m}\mathcal M_{j,m}$, one has $\mathcal M_m^{*}=\sigma_m\sum_{j\in\mathcal K_m}q_{j,m}$. If the aggregate is non-zero, then $\sign(\mathcal M_m^{*})=\sigma_m$, the source-based sign then agrees exactly with the primary sign. 

%The flux-based sign requires extra monotonicity, local response conditions such that $\sigma_m \Delta F_m>0$, and forcing-dominance assumptions, and is therefore only conditionally equivalent. 
If source-based or flux-based signs are noisy observations of the primary sign, the LMF exponent is preserved only under sufficiently weak or short-memory sign misclassification noise. If misclassification is correlated with meta-order length or liquidity state, the measured exponent can differ. This means that long-memory is stable under sign noise but not arbitrary transformations. Consider for example noisy sign measurements $\tilde \epsilon_m = \eta_m \epsilon_m$ for independent sign noise $\eta_m = \{+1,-1\}$. Then $\E[\tilde \epsilon_m \tilde \epsilon_{m+\tau}] \propto \E[\epsilon_m \epsilon_{m+\tau}]$; if the sign transformation is state-dependent, clustered, or correlated with liquidity and meta-order length, then the exponent may change.

\subsection{Renewal derivation of the LMF law} \label{sec:lmf}

% cite Sato & Kanazawa (2024) not only for support but also for caveats: that the exponent is robust but the prefactor is heterogeneous-strategy-sensitive.

The child-order flow is modelled as sequence of non-overlapping meta-orders with i.i.d. signs and i.i.d. lengths \cite{sato2023prl,sato2024jsp}. The meta-orders are indexed by $n=1,2,3,\ldots$ where the $n$-th meta-order is characterised by the order sign $\sigma^{(n)}_m \in \{ +1 , -1 \}$ and has length $L_n \in \N$ in the child-order event count at event $m$. In reality there is meta-order mixing as multiple meta-orders arrive at the market. 

Hence we need to assume that if two events are not in the same meta-order, the expected product of their signs is zero only because the meta-order signs are assumed independent and symmetric {\it i.e.} $\E[\sigma^{(n)}]=0$ and $\E[\sigma^{(n)} \sigma^{(n')}] =0$ for $n \neq n'$ so there are no correlations across meta-orders.

Here a single meta-order arrives and it is incrementally executed. The meta-orders lengths are i.i.d. integer-valued random variables with the law: $p_L(\ell) = \Pp[L= \ell]$ where the mean length is finite $\E[L] = \sum_{\ell=1}^\infty \ell p_L(\ell) < \infty$. The resulting event-time sign process is thus $\epsilon_1,\epsilon_2,\epsilon_3,\ldots$ where a given $\epsilon_m$ is constant over the lifetime of a given meta-order, and changes sign only when a new meta-order begins.

Let $\widetilde p_L(\ell)$ denote the length-biased law seen from a uniformly sampled child-order event
\begin{equation}
\widetilde p_L(\ell)=\frac{\ell p_L(\ell)}{\mathbb E[L]}.
\label{eq:ellbiasedlaw}
\end{equation}
Here a meta-order of length $\ell$ contributes exactly $\ell$ child-order events to the stationary sequence. This means that relative to the original meta-order law $p_L(\ell)$ the probability of being observed at a uniformly sampled event is proportional to $\ell p_L(\ell)$; this is then normalised by the total expected number of child-order events per meta-order.

For a fixed lag $\tau$ we can condition on the event that the sampled child-order belongs to a meta-order of length $\ell$. The lagged event remains in the same meta-order iff the sampled position $r$ satisfies $r+\tau\le \ell$. Hence the overlap probability is
\begin{equation}
\mathbb P[\mbox{same meta-order} \mid L=\ell]=
\begin{cases}
(\ell-\tau)/\ell,&\ell>\tau,\\
0,&\ell\le\tau.
\end{cases} \label{eq:condsign}
\end{equation}
Multiplying by the length-biased law and summing over $\ell$ yields the exact renewal formula.

Under the sequential meta-order assumption the sign autocorrelation is then 
\begin{equation}
    C_\tau(\epsilon) = \sum_{\ell=1}^{\infty} \E[\epsilon_m \epsilon_{m+\tau} | L =\ell] \tilde p_L(\ell).
\end{equation}
Since the sign remains constant within the meta-order, from the conditional probability of sign agreement at lag $\tau$ given by equation \ref{eq:condsign} the conditional auto-correlation reduces to
\begin{equation}
    C_\tau(\epsilon) = \sum_{\ell=\tau+1}^{\infty} \frac{(\ell-\tau)}{\ell} \tilde p_L(\ell).
\end{equation}

From the length-biased law in equation \ref{eq:ellbiasedlaw} we then find equivalently that the event-time sign autocorrelation satisfies the renewal formula 
\begin{equation}
    C_\tau (\epsilon)=\frac{1}{\mathbb E[L]}\sum_{\ell=\tau+1}^{\infty}(\ell-\tau)p_L(\ell).
\end{equation}
If the meta-order length distribution has Pareto tail $p_L(\ell)\sim c_L\ell^{-(\alpha+1)}$ with $\alpha>1$, then by approximating the sum as an integral and evaluating the integral\footnote{
$\sum_{\ell=\tau+1}^{\infty}(\ell-\tau)\ell^{-(\alpha+1)}
\sim
\int_\tau^{\infty}(\ell-\tau)\ell^{-(\alpha+1)}d\ell
=
\frac{1}{\alpha(\alpha-1)}\tau^{1-\alpha}$}
\begin{equation}
    C_\tau(\epsilon)\sim \frac{c_L}{\alpha(\alpha-1)\mathbb E[L]}\tau^{-(\alpha-1)},
\end{equation}
If the event-time trade-sign auto-correlation satisfies $C_{\tau}(\epsilon) \sim K \tau^{-\gamma}$ this then yields the LMF exponent relation $\gamma=\alpha-1$ \cite{lillo2005theory,sato2023prl,sato2024jsp}. This is the precise statement that the heavy tailed meta-order lengths generate long-memory in trade-signs. This mechanism is renewal-theoretic because long meta-orders occupy many child-order event slots. This means that event-time sampling overweights long meta-orders via the length-biased law. Then large lags are dominated by the probability that two events still belong to the same unusually long meta-order. 

The connection back to the reaction-diffusion model is because the renewal equation is expressed entirely in terms of primary sign process $\epsilon_m =\sigma_m$. If the source-based sign agrees with the primary sign under the signed localised forcing, and if the flux-based sign agrees conditionally in the forcing dominant regime, then within the reaction-diffusion interpretation, then the LMF law is a statement about the persistence of the effective interface forcing (or the interface response) to long meta-orders. 

\subsection{Weak return autocorrelations} \label{sec:efficiency}

% 1. lillo2004efficient for long-memory order signs plus liquidity/size compensation, 
% 2. Bouchaud et al. (2004) and Bouchaud et al. (2006) for transient propagators / liquidity molasses / critical response,
% 3.  farmer2013impact for fair pricing / efficiency-shaped impact arguments

%  Bouchaud et al. (2004), Bouchaud et al. (2006), Lillo & Farmer (2004), and Farmer et al. (2013) need to be more carefully cited.

The event-time price fluctuations are the return $r_m$ in terms of the differences between log-prices that define the reaction front: $r_m = p_m - p_{m-1}$ where $p_m$ is the front (mid-price) sampled at child-order event time $m$. The return autocorrelation is then
\begin{equation}
    C_\tau(r) = \Cov[r_m,r_{m+\tau}], \quad \tau \in \N. \label{eq:returncov}
\end{equation}
In sufficiently liquid markets a standard measure of market efficiency is that this autocorrelation is zero or at least very small for $\tau>0$, despite the sign autocorrelation $C_\tau(\epsilon)$ decaying slowly. 

The exact front equations are in local time but depend on the full field $\phi(x,t)$. After linearisation around a stationary background and integration of the perturbation dynamics, the displacement front becomes a weighted superposition of past signed order flow. This motivates a propagator representation \cite{bouchaud2018tqp}. A linearised propagator representation consistent with the Green-function picture is
\begin{equation}
    r_m\approx \sum_{\ell\ge 0}G_\ell\epsilon_{m-\ell} q_{m-\ell}+\xi_m, \label{eq:returnprop}
\end{equation}
Here $G_{\ell}$ is the effective propagator at event lag $\ell$, $\xi_m$ is an orthogonal innovation representing noise not explained by the linear component. The effective propagator is the event-time footprint of the diffusion/replenishment kernel after discretisation and local linearisation of the front. Some care is required here because two things are happening the first is a discrete sampling, in the physical time (assuming it is mapped into the continuum operational time of the front), and second, that the sign of the order schedule at the sampled front can be identified with the primary trade sign. This is exact in the original discrete event time formulation it is not going back from the continuum representation by sampling.  

This is not an exact reformulation of the full non-linear front dynamics. Rather, it is the event-time (more correctly the operational time sampling) linear-response approximation induced by the Green-function picture --- it isolates the mechanism by which persistent signs can be filtered into weakly autocorrelated returns. The return covariance becomes a double convolution of the propagator with the sign-memory kernel. Persistent sign memory therefore need not imply persistent return memory: diffusion, replenishment, front-slope adaptation, and state dependence can all suppress return predictability even when order flow has long-memory. 

From the return autocovariance in equation \ref{eq:returncov} if we assume that the event-time propagator returns have zero mean and orthogonal innovations then the linearised propagator is
\begin{align}
C_{\tau}(r)
&\approx
\sum_{\ell\ge0}\sum_{k\ge0}G_\ell G_k\,
\Cov[\epsilon_{m-\ell}q_{m-\ell},\epsilon_{m+\tau-k}q_{m+\tau-k}] \nonumber \\
&\quad +
\Cov(\xi_m,\xi_{m+\tau}).
\end{align}
Under leading-order signed-volume factorisation 
\begin{equation}
\Cov[\epsilon_{m-\ell}q_{m-\ell},\epsilon_{m+\tau-k}q_{m+\tau-k}]
\approx \bar q^{\,2}C_{\tau+\ell-k}(\epsilon),
\end{equation}
where $\bar q$ is the average trade size. Return covariance is then a double convolution of the propagator with sign memory
\begin{equation}
    C_{\tau}(r) \approx \bar q^2 \sum_{\ell\ge0}\sum_{k\ge0}G_\ell G_k\,
C_{\tau+\ell-k}(\epsilon) + C_{\tau} (\xi).
\end{equation}
If the propagator decays as: $G_{\ell} \sim \ell^{-\beta}$ \cite{toth2011anomalous} for large $\ell$ then convolution is substantially more transient than the raw sign process whenever the propagator decay is strong. There are broadly two mechanisms by which this filtering can suppress return predictability: transient impact \cite{donier2015fully}, and state-dependent liquidity adaption. 

Transient impact: if the propagator decays sufficiently quickly in the lag, then the contribution of a predictable continuation to future returns is small; persistent order signs move the price transiently, but leave only a weak long-lag return covariance. 

Liquidity adaption: in the reaction-diffusion picture, the incremental price response to predictable order flow can be reduced because liquidity replenishment and front-slope adaption respond directly to the persistence of execution. This means that the effective propagator is not fixed once and for all, but rather depends on the local book state. This makes the response weaker than a naive fixed-impact model could predict. 

In summary: in the reaction-diffusion picture there are several reasons why this cancellation can occur. First, the front velocity depends on the inverse local slope of the order book rather than on a fixed and permanent impact parameter. Second, the perturbation generated by each child order is diffused over time (through the Green kernel). Third, the stationary background sources, both lit and latent, replenish liquidity and oppose unrestricted directional drift. Finally, the meta-order persistence that generates long sign memory tends to modify the local state of the order book in a way that changes the incremental impact of future child orders. Together these imply that a persistent sign sequence need not produce a persistent return sequence. 

This is one of the successes of the reaction-diffusion model for high-frequency finance because when we derive the infinitesimal local response formula for small changes in the front $\delta p \approx \delta q / {\cal L}$ we see clearly that local linearity does not imply predictability. The local response formula is a one-step static relation evaluated at a given front slope; efficiency, in the sense of weak return autocorrelations, depends instead on the cumulative dynamic response after diffusion, replenishment, and slope adaption have acted on the sequence of child orders. This explains why in the model local linearity and weak return autocorrelation are compatible. What remains unclear is what depends on the event ordering from what depends on the mappings between event time and the physical clock time. 

\subsection{Subordination} \label{sec:subordination}

% need it include Montroll–Weiss and Metzler–Klafter

Event time is the discrete ordering of child-order events $m=1,2,3,\ldots$; all presistence statements about the sign sequence $\epsilon_m$ in the LMF renewal derivation are statement in this event-time. The calendar-time of physical clock times of child-order events is obtained from the cumulative waiting times between the child-order events
\begin{equation}
T_m=\sum_{k=1}^{m}\Delta t_k,
\qquad T_0=0,
\end{equation}
The inverse event counter then gives the number $N_t$ of child-order events that have occurred by clock time $t$
\begin{equation}
N_t=\max\{m\ge0:T_m\le t\}.
\end{equation}
Then the physical-time front process is the subordinated event-time process
\begin{equation}
X(t)=Y_{N_t},\qquad Y_m:=p_m.
\end{equation}
Here the event-time process $Y_m$ carries the sequence level dynamics of the order flow and front updates, while the counting process $N_t$ maps those dynamics onto the physical clock {\it i.e.} the market evolve by event count (in operational time) first, and clock time (calendar time) is obtained by forcing a random waiting-time accumulation. If the waiting times have finite mean $\bar{\Delta t}$, then $T_m\sim m\bar{\Delta t}$ and $N_t\sim t/\bar{\Delta t}$ for large times. Hence an event-time law $Y_m\propto \sqrt{m}$ maps to a clock-time law $X(t)\propto \sqrt{t}$ up to a deterministic rescaling. 

If the waiting times are heavy-tailed, then $N_t$ is no longer asymptotically proportional to $t$, so clock-time impact can be anomalous even though the event-time LMF law remains unchanged. Why this is important is because the Volterra equation and square-root impact laws are statements about the delivery of signed flow over physical time; in particular, the execution rate $m(t)$  and the completion horizon $T$ are clock time artifacts of moving from the DTRW to a CTRW. This is clear because the forcing is in fact written as a physical time source $m(t) \delta(x-p(t))$, the Green function is a function of the clock-time lag $(t-s)$, and the execution horizon is the physical interval $[0,T]$. These all depend explicitly on the uniqueness of time change from event count to clock time. 

This means that two meta-orders with the same event-time sign sequence can have different physical-time impact paths if their waiting-time sequences differ. Consider, for example, the signed volume $Q_m = \sum_{k=1}^m {\epsilon_k q_k}$ where $q_k \ge 0$ is the size of the $k$-th child order. Then the corresponding physical-time cumulative signed volume is $Q(t) = Q_{N_t}$. If the waiting times are approximately constant then event count and clock time are approximately proportional. However, if waiting times are strongly heterogeneous then this proportionality is broken and the same event-time path becomes stretched or compressed in physical time. 

Finite mean waiting times allow one to argue for an asymptotic proportionality of event-time and physical-time, but some care is prudent. Even though at leading order event count and physical time differ, this is then only by a constant rescaling --- it is only in this regime that event-time and clock-time descriptions can agree (up to a change in units) and hence only in this regime that the leading square-root exponent is unchanged. 

This implies that for heavy-tailed waiting times the cumulative clock process will grow anomalously and then the inverse counter $N_t$ is no longer proportional to physical time $t$. In this more realistic regime the event-time process may still have LMF long-memory (because the law depends only on the event ordering). However, the physical-time impact trajectory $X(t) = Y_{N_t}$ can exhibit anomalous clock-time scaling because the event counter itself can evolve sublinearly in $t$. This will modify the physical-time transient impact profile, the clock-time duration of the meta-order (of a fixed event-time length), and the mapping between the event-time and physical-time autocorrelations.

\section{Discussion}
\label{sec:discussion}

The main contribution of the paper is the derivation of two established laws---the square-root impact law and the LMF long-memory law---within one common discrete bid/ask order-book model. In the existing literature these two phenomena are usually explained in different languages: reaction--diffusion or latent-order-book theory for impact \citep{toth2011anomalous,donier2015fully,benzaquen2018multi,bouchaud2018tqp}, and sequential hidden-order or renewal arguments for sign persistence \citep{lillo2005theory,sato2023prl,sato2024jsp}. We show how both laws can be generated from the same signed forcing structure in event time once the order book is written discretely at the bid/ask level and then reduced to the discrete imbalance field.

The discrete imbalance reduction is the organising bridge. The use of a signed liquidity or latent imbalance field is not new in itself; it is already natural in the latent-liquidity and reaction--diffusion literature \citep{toth2011anomalous,donier2015fully,mastromatteo2014latent}. What is distinctive here is that the exact discrete cancellation of the local matching term is made the central bridge between the microscopic bid/ask model and the mesoscopic front-dynamics description. Here this exact discrete identity comes before the formal continuum limit and therefore cleanly identifies which simplification is exact and which are merely continuum approximations. The continuum derivations can be considered conventional, but repeated to highlight specific features important to our line of reasoning with focusses on the time mixing and how the forcing terms are most naturally formulated. 

We also try to be explicit in terms of moving between order-splitting and impact models: the difference between the meta-order sign itself, the sign of the forcing entering the imbalance equation near the front, and the sign of the event-induced change in front flux. The LMF literature focuses on the sign process as such \citep{lillo2005theory,sato2023prl,sato2024jsp}, whereas the reaction--diffusion and LLOB literatures focus on forcing, front motion, and impact \citep{donier2015fully,benzaquen2018multi,bouchaud2018tqp}. By defining and comparing these three notions of sign explicitly we are trying to bridge the two literatures.

Event time, physical time, and the DTRW perspective are central to our approach. We think that the explicit separation between event time, operational time and physical time, and the use of a DTRW/subordination interpretation to organise these distinctions is useful. The broad stochastic-process idea is not new: DTRW, CTRW, and subordination are established tools in anomalous diffusion and related transport models \citep{angstmann2015dtrw,meerschaert2012fractional,diana2024anomalous}. What we think is new here is the use of this perspective to make explicit that the LMF law is fundamentally an event-time statement, whereas the square-root law is fundamentally a physical-time statement. This provides an explanation of why non-uniform waiting times do not alter the event-time sign-memory law but can alter the clock-time impact trajectory.

In summary there are three distinct time variables in the construction.  First, there is the
event time $m\in\mathbb N$, which orders child-order arrivals and is the natural
time of the trade-sign sequence $\epsilon_m$.  Second, there is the operational
time (say) $u$, obtained when the discrete order-book dynamics are passed to a continuum
reaction--diffusion representation.  The Volterra equation for the moving front is in fact
derived in this operational time.  Third, there is the physical calendar time
$t$, in which empirical meta-order impact, execution duration, and participation rates
are usually measured.  The standard square-root law is a statement about impact at
calendar completion, not a statement about the event-time sign process.

This three-clock distinction leads to two different anomalous-time interpretations.
The first is the inverse-subordination or event-driven interpretation:
events move the book and the price; the calendar clock only randomises when those
events occur.  In that case the operational/event-time impact law is unchanged. The second interpretation is a physical-time fractional transport model:
liquidity transport itself is anomalously slow in calendar time.  Then the continuum
order-book equation is not merely time-changed after the fact; its physical-time limit
is fractional. 

Here we have also tried to maintain an explicit hierarchy between exact results, formal continuum limits, and controlled leading-order approximations. The exact discrete imbalance reduction, the exact front differentiation identity, and the exact renewal formula are separated from the formal diffusion limit, the full-line Green-function idealisation, and the small-displacement/weak cancellation reduction used to derive the square-root law. The market-impact literature often mixes these logical layers because empirical law, no-arbitrage arguments, and continuum intuition are discussed together \citep{gatheral2010nda,donier2015fully,benzaquen2018multi}. 

\section{Conclusion}
\label{sec:conclusion}

The empirical existence of long-memory order flow, the order-splitting explanation of the LMF exponent relation, and the latent-order-book explanation of square-root impact are all established \citep{lillo2005theory,toth2011anomalous,donier2015fully,sato2023prl,sato2024jsp,benzaquen2018multi}. 

The central message here is that the long-memory law and the square-root law live on different clocks. The trade-sign law is an event-time statement: it concerns the ordering of child orders and the persistence created when long hidden orders are split into many same-sign pieces \cite{lillo2005theory,sato2023prl}. The impact law is an operational-time statement: it concerns the evolution of an imbalance field and its moving reaction front under diffusive transport, replenishment, and signed forcing \cite{toth2011anomalous,donier2015fully}. Calendar-time observables are then obtained by subordination through the waiting-time process between order-book events, in direct analogy with discrete-time random walk constructions used in anomalous diffusion \cite{angstmann2015dtrw,meerschaert2012fractional,diana2024anomalous}.

\section{Acknowledgements}

Derick Diana, Byron Jacobs and Dominic Bauer. Copilot was used to help with the editing and some of the reference checking, in particular ChatGPT 5.5 was used to check some of the results and to help with some editing.  
%% BIBTEX
%\bibliographystyle{IEEEtran}
\bibliography{RD2LMF-v2}
% \appendix

\end{document}